\documentclass[twocolumn,prl]{revtex4}
\usepackage{times}
\usepackage{amsmath}

\usepackage{graphicx}

\usepackage{color}
\definecolor{red}{rgb}{1,0,0}
\definecolor{blue}{rgb}{0,0,1}

\renewcommand{\thefigure}{\textbf{\arabic{figure}}}

\begin{document}

%\title{Simulated Chromosome-wide Dynamics: two-step chromatin folding and self-assembly of nuclear bodies}
\title{Chromosome-wide simulations uncover folding pathway and 3D organization of interphase chromosomes}

%\tightenlines

\author{Davide Michieletto$^1$, Davide Marenduzzo$^1$$^*$, Ajazul H. Wani$^2$$^*$}
\affiliation{$^1$ SUPA, School of Physics and Astronomy, University of 
Edinburgh, Peter Guthrie Tait Road, Edinburgh, EH9 3FD, UK\\$^2$ Department of Biotechnology, University of Kashmir, Srinagar, J\&K, India, 190006. \\$^*$ Corresponding authors: dmarendu@ph.ed.ac.uk, ahwani@kashmiruniversity.ac.in}

\begin{abstract}
\textbf{Three-dimensional interphase organization of metazoan genomes has been linked to cellular identity. However, the principles governing 3D interphase genome architecture and its faithful transmission through disruptive events of cell-cycle, like mitosis, are not fully understood. By using Brownian dynamics simulations of \emph{Drosophila} chromosome 3R up to time-scales of minutes, we show that chromatin binding profile of Polycomb-repressive-complex-1 robustly predicts a sub-set of topologically associated domains (TADs), and inclusion of other factors recapitulates the profile of all TADs, as observed experimentally. Our simulations show that chromosome 3R attains interphase organization from mitotic state by a two-step process in which formation of local TADs is followed by long-range interactions. Our model also explains statistical features and tracks the assembly kinetics of polycomb subnuclear clusters. In conclusion, our approach can be used to predict structural and kinetic features of 3D chromosome folding and its associated proteins in biological relevant genomic and time scales.
}	

\end{abstract}

\maketitle

\section{Introduction}

%Introduction
Eukaryotic genomes have a sophisticated hierarchical three-dimensional organization, ranging from nucleosomes to loops, topologically-associating domains (TADs)~\cite{Lieberman-Aiden2009,Sexton2012,Dixon2012,Rao2014} and chromosome territories~\cite{Cremer2001}. Understanding this multi-scale organization is a fundamental problem in biology as it has been linked to development, differentiation and diseases~\cite{Bickmore2013,Cavalli2013,Zane2014}. Application of chromosome conformation capture (3C) and microscopy-based methods has provided unprecedented insight into this problem. 3C-based methods have shown that genomes of metazoans are organized into TADs separated by boundaries~\cite{Lieberman-Aiden2009,Sexton2012,Dixon2012,Rao2014}. While the mechanism leading to the formation of TADs and the biological nature of boundaries are still not fully understood, good correlations between enrichment of specific proteins and histone modifications in different TADs have been observed~\cite{Sexton2012,Dixon2012}. 

Chromatin-binding proteins, usually present in the form of multiprotein complexes, interact through a combination of protein-protein and DNA-protein interactions and can provide the enthalpy to compensate for the entropic costs associated with the formation of TADs or other compacted or looped forms of chromatin. The multiprotein complex, Polycomb repressive complex 1 (PRC1), can compact chromatin {\it in vitro}, and it is known that {\it in vivo} it bridges chromatin regions separated by tens of kbs to Mbs~\cite{Wani2016}. The ability of PRC1 to mediate intra-domain as well as long-range chromatin interactions stems from its self-association property mediated by SAM-domain polymerization of its PH subunit~\cite{Wani2016}. Heterochromatin protein 1 (HP1) also self interacts and stabilizes heterochromatic regions. Although CTCF and other insulator proteins mark boundary regions, they are also found within TADs and can mediate long-range loops either by homotypic or heterotypic protein-protein interactions~\cite{Seitan2013}.

Association of chromatin-bound proteins with each other also regulates their subnuclear distribution. Microscopy studies have uncovered subnuclear organization of some proteins into clusters, or \emph{foci}. Polycomb group (PcG) proteins are organized into hundreds of nanoscale clusters and their subnuclear distribution is coupled to chromatin topology~\cite{Wani2016}. In addition, ``transcription factories'' underlying active gene clusters, Cajal bodies and nuclear speckles are other features of subnuclear protein distribution linked to gene expression~\cite{Sleeman2014,Papantonis2013}. Super-resolution microscopy coupled to FISH, in contrast to Hi-C, a population averaged technique, can provide single cell information and physical scales of various chromatin features~\cite{Kritikou2005,Spector2004,Boettinger2016}.

Many of the chromatin-associated proteins regulating interphase organization dissociate from chromatin when cells enter into mitosis, where chromatin is reconfigured~\cite{Blobel2009,Fanti2008,Zhao2011,Buchenau1998,Xing2008,Fonseca2012,Follmer2012}. As a consequence, how interphase chromatin features are preserved through chromatin reconfiguration stages of cell division is not well understood. Faithful inheritance and imprinting of epigenetic traits regulating interphase chromatin organization from parent cell to daughter cells ensures maintenance of proper gene expression which, in turn, determines cellular identity~\cite{Steffen2014}. Monitoring real time relaxation of chromosomes from the mitotic to the interphase state would shed much light into this complex issue, but such experiment has not been realised yet, as it appears very challenging and demands extensive high resolution observations.\\

In parallel with this rapidly growing body of experimental data, there have been a number of polymer physics models aiming at elucidating different aspects of genome organization within eukaryotic nuclei. Some of the models~\cite{Nora2012,Bau2012,Zhang2015} start from the Hi-C data and have the goal of finding an appropriate force field for the intra-polymer interactions which can recreate these data. A complementary approach, so called ``bottom-up'', starts from basic biological assumptions and on the sequencing, \emph{e.g.}, chromatin immuneprecipitation coupled to sequencing (ChIP-Seq), datasets characterizing the one-dimensional (1D) genomic features. From this information, but rigorously avoiding any prior knowledge of three-dimensional (3D) contacts or genomic architecture, these models explore via large-scale computer simulations the resulting 3D spatial organization of the genome~\cite{Barbieri2012,Brackley2013,Jost2014,Benedetti2014,Brackley2015,Brackley2016}. In this case, the predicted 3D structure can then be directly compared with Hi-C and microscopy data. The two approaches are also sometimes denoted as ``inverse'' and ``direct'' modelling respectively. Our current work is an example of the ``direct'' modelling avenue, and its key strength is in its much enhanced predicted power. As our direct modelling approach does not involve computationally costly fitting procedures, we can simulate very large regions, in this case entire chromosomes, and follow their dynamics for time-scales which are biologically relevant, up to several minutes. These two aspects of our approach enable us to make prediction of structural and kinetic features of chromosome relaxation in biologically relevant genomic and time scales.

In this work we use Brownian dynamics simulations to study the folding of the entire {\it Drosophila}  chromosome 3R (chr3R). In our model, this chromosome is viewed as a semi-flexible polymer made up of coarse grained beads (each representing about 3 kbp of the fly's genome, see SI, Methods), which interact with soluble proteins, or protein complexes. Our minimal model is based on the simple biological assumption that protein complexes are mainly responsible for driving the observed 3D organization of the chromatin~\cite{Nicodemi2009,Barbieri2012,Brackley2013,Johnson2015,Cheng2015}. Since these proteins can bind to chromatin at many places, they can bridge non-contiguous chromatin regions to which they are bound. 
Loops mediated by CTCF, chromatin contacts mediated by polymerization of PcG protein PH~\cite{Beisel2011}, and heterochromatin-associated complexes such as dimers of HP1~\cite{Kilic2015} are some examples of chromatin contacts orchestrated by a combination of DNA-protein and protein-protein interactions. 

For simplicity, we only directly model protein bridges, and disregard monovalent proteins. The affinity of chromatin beads for a protein complex is either directly set according to the chromatin immunoprecipitation (ChIP-seq) data, or as per the local chromatin states, which have previously been inferred from various ChIP-seq data~\cite{Filion2010,Karchenko2011}.

In order to test the predictions of our model, we compare our simulations with both, experimentally obtained chromatin contact maps and subnuclear size distribution of protein clusters. In all cases, we find good overall agreement between simulations and experimental observations, which is remarkable given the simplicity of the assumptions in our model. In addition, we also track the folding dynamics of the chromosome up to experimentally relevant time-scales of several minutes. From this study we draw an important conclusion: the folding dynamics of an interphase chromosome proceeds in a two-step fashion. First, local contacts within a TAD form fast and reproducibly; concomitantly, nuclear protein clusters self-assemble within minutes. Later on, non-local contacts between genomic regions in different TADs are established. This process is slower and it is accompanied by an equally slow coarsening of the nuclear protein clusters. 
Importantly, the first step (local folding into TADs and nuclear protein cluster assembly) appears solely dependent on the protein binding landscape along the chromosome and strongly suggests that this minimal 1D information can lead to the faithful 3D organization of the genome. On the other hand, there is a notable dependence on the chosen initial chromosomal state in the higher order folding; the formation of non-local, inter-TAD interactions (which is accompanied by coarsening of nuclear protein clusters) has also a significant stochastic component, as different contacts are set up in different simulations -- similarly to what observed in single cell Hi-C experiments. In what follows, we will refer to both dependencies on initial conditions and on stochasticity by saying that the second folding step depends on the conformational ``history'' of the system.

\section{Results}

\begin{figure*}[t]
	\includegraphics[width=1.0\textwidth]{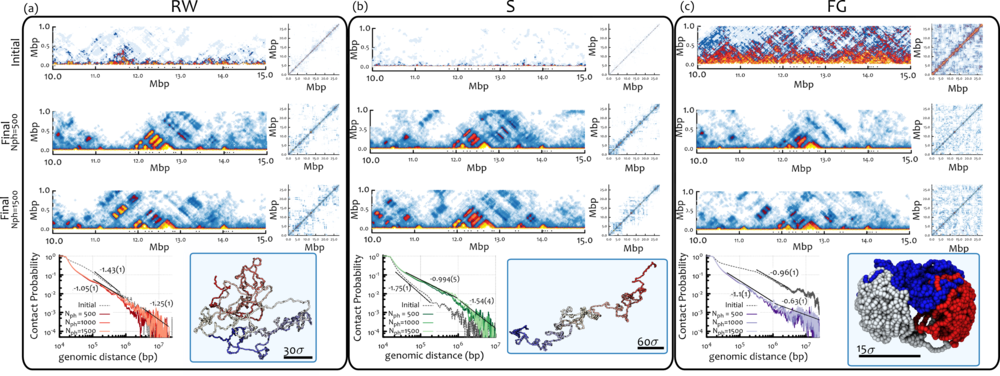}  %{../figures_3/fig1_plusFISH.eps}
	\caption{\textbf{PcG protein driven topology of chr3R in simulations: Local domains form robustly and similarly for all initial chromosomal states, whereas non-local contact patterns reflects the system history.} Here we show the contact maps and the contact probability for different initial configurations and concentration of binding proteins. The top three rows show contact maps: the left picture shows the contact map which is rotated by 45 degrees and truncated to show more clearly local domains along the diagonal in the region 10-15 Mbp; the right picture shows the full contact map. The three rows correspond to, respectively: starting configuration (top); final configuration with 500 PRC1 proteins (middle); final configuration with 1500 PRC1 proteins (bottom). The final configurations are obtained after $2$ $10^5$ simulation time units, or 2-20 minutes (see SI for the mapping between simulation and physical units). The fourth row shows the decay of the contact probability as a function of genomic distance in the initial (grey line) and final configurations (coloured lines), and a snapshot of the initial configuration (right picture). In the contact probability plot we also show power law fits with different effective exponents in the intradomain ($<$Mbp) and interdomain ($>$Mbp) range. \textbf{(a)} \textbf{(b)} and \textbf{(c)} correspond to Random walk (RW), stretched (S) and fractal globule (FG) initial chromosomal states. See text and SI for further details on the preparation. By comparing initial and final rows one can notice that local domains form robustly and similarly for all initial conditions, whereas non-local contact patterns depend much more strongly on initial conditions. This is also supported by the two exponents found for the decay of contact probability: close to -1 for sub-megabase contacts and for all cases, while case-subjective for the inter-TAD ($>$Mbp) regime. See also Suppl. Movies M1-M3.
	}
	\label{fig:Domains}
\end{figure*}

\subsection{PRC1 binding pattern robustly predicts the formation of a sub-set of the experimentally observed TADs}

We begin by modelling the chromatin fibre of \emph{Drosophila} chromosome 3R (chr3R) during interphase as a semi-flexible polymer (see SI, Methods) interacting with soluble proteins, representing the polycomb repressive complex 1 (PRC1). This complex binds to chromatin at many places, and a subunit of it, polyhomeotic PH,  self-polymerizes, making PRC1 able to effectively bridge between different regions of the chromatin fiber~\cite{Wani2016}. The affinity between PRC1 and each chromatin bead is informed by ChIP-seq data (see SI, Methods) and can vary between two possible values: high affinity binding (interaction strength $\epsilon_1$) and weaker affinity (interaction strength $\epsilon_2 < \epsilon_1$). The relative affinities of these two types of binding sites are set accordingly to the relative binding intensities obtained from ChIP-seq data~\cite{Wani2016} (see SI Table~T1 and Fig.~S1).
 
We initialized the simulations from three possible conformations of the polymer thereby simulating different initial chromosomal states: (i) random walk (RW), (ii) stretched random walk (S) and (iii) fractal globule (FG) (see right-bottom corner of Figs.~\ref{fig:Domains}(a),(b) and (c), respectively, for examples of such configurations). While the random walk represents the simplest polymer physics choice (see Suppl. Movie M1), the stretched random walk may better emulate a post-mitotic elongated fiber (Suppl. Movie M2). Finally, the fractal globule state is motivated by the homonymous model, according to which interphase chromosomes are folded so that they can crumple and segregate whilst remaining entanglement-free. As a consequence of this peculiar global folding, contacts in the fractal globule are established almost exclusively among genomic regions which are also contiguous in 1D sequence space~\cite{Mirny2011} (see Suppl. Movie M3). %%DM added for point 2) Knottiness
\textcolor{black}{Although all initial configurations are prepared in such a way that they possess a small knotting probability, we do not check for the presence of knots in either initial or final configurations. Nonetheless, the chain is allowed to cross through itself with an exponentially small probability (see SI) in order to quickly move away from highly entangled, topologically frustrated, configurations.}

Starting from these initial configurations, we followed the evolution of contacts of  chr3R for about 2-20 minutes of real time (calculated assuming a nucleoplasmic viscosity of 10-100 cP, respectively, ten or a hundred times that of water, see SI, Methods). Remarkably, we find a very similar local domain structure for all three initial conditions (as shown in the zoom of the region 10-15 Mbp in the second and third rows of Fig.~\ref{fig:Domains}(a)-(c)), which suggests that the formation of some topological domains is largely driven by PRC1 binding and self-association. 

Interestingly, in the case of the fractal globule, while at the beginning the domains of enriched contacts are visible at multiple scales (see top row in Fig.~\ref{fig:Domains}(c)), as expected for a fractal structure, later on these domains rearrange to settle into their final locations, which is determined by the ChIP-seq binding sites of PRC1, and bear little resemblance with the initial state (see Fig.~\ref{fig:Domains}(c) and compare second and third rows with first row, see also SI). Remarkably, the final domain locations predicted from all three different starting conditions are in good agreement with TADs as found in Hi-C experiments~\cite{Sexton2012} and classified as PcG TADs (see SI Fig.~S2).

Differently to the local domain folding, the non-local pattern of contacts, visible as the off-diagonal structure in the full contact map (reported in Fig.~\ref{fig:Domains} for the various cases), is more strongly affected by the starting configuration -- the contact maps also show variability in different runs, so that several runs need to be averaged when computing the contact map. For the stretched random walk, local interactions ($<$1 Mb away along the genomic sequence) account for well over 90\% of the overall contacts. This number falls to about 80\% for the RW initial condition, and is similar for the fractal globule conformation. An analysis of the contact probability versus genomic distance ($P(s)$,  Fig.~\ref{fig:Domains} bottom row) confirms that there is a ``local'' scale where the information regarding the initial chromosomal state is lost and a ``non-local'' scale which carries the signature of system history. This can be seen from the power law decay exponent of $P(s)$. Interestingly, we find that restricting the datasets to contacts between regions closer than 1 Mbp in sequence, \emph{i.e.}, ``local'' interactions, the effective exponent measuring the decay of contact probability with distance (within chr3R) is close to -1 whatever the initial chromosomal state. 
%(the average between the 3 initial conditions considered is -1.05). 
%REMEMBER THAT HERE WE ARE USING ONLY PcG -- CANNOT COMPARE DIRECTLY WITH EXP
%USE INSTEAD FULL MODEL
This values becomes closer to the experimentally observed one~\cite{Sexton2012} ($\sim -0.75$) when a more detailed model is considered (see later sections and SI Fig.~S9). 
Nonetheless, this history-independent emerging exponent reflects the fact that the local organization is solely determined by chromatin-associated proteins and their binding profiles, rather than the initial chromosomal architecture, which, as we showed, is lost in steady state (Fig.~\ref{fig:Domains}). The exponent characterizing non-local interactions is instead observed to vary significantly with initial condition, which once again suggests that the inter-TAD organization is dependent on the system history.
% with the stretched configuration leading to a conformation displaying an effective exponent around -1.54, and the fractal globule one leading to a much shallow decay of non-local contacts, with an effective exponent of -0.63. The near-equality of -1.54 with -1.5, the exponent expected for contacts in a simple random walk, is likely only a coincidence, as the RW start leads to a different steady-state non-local exponent (-1.25). 
The fact that there can exist two effective exponents: one for local, intradomain interactions, and another one for non-local, or interdomain, interactions is also supported by the most recent HiC data for contacts within human chromosomes~\cite{LibAid2015}. These findings also complement the ones in Ref.~\cite{Barbieri2012}, which demonstrated that the effective exponent is dependent on interaction strength and protein concentration. In our simulations we in fact observe that the effective exponent for long-range contacts depends on the history of the system, which translates into a sensitivity on timing of the cell cycle or on sample preparation.

\begin{figure*}[t]
	\includegraphics[width=0.99\textwidth]{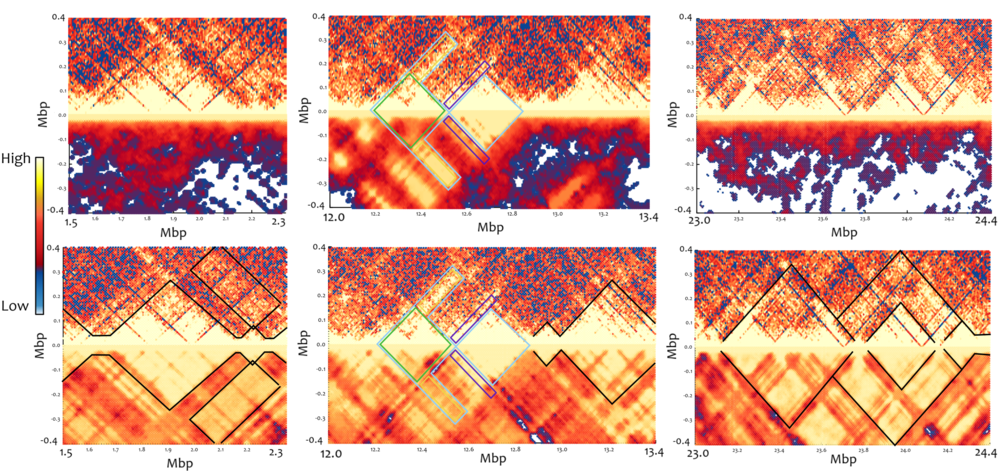}  
	\caption{\textbf{Comparison between predicted and experimentally observed chromatin topology using ``polycomb-only'' or ``full'' model simulations.} In each of these plots we rotate the contact maps in order to better compare the TADs obtained from simulations (bottom halves) to the ones from Hi-C experiments~\cite{Sexton2012} (top halves) for three chosen segments of ch3R (left:1.5-2.35Mbp; middle:12-13.4Mbp; right:23-24.4Mbp). The top row shows the comparison with the ``polycomb-only'' model while the bottom row with the ``full'' model. It is evident that the latter better captures the position and the fine structure of the experimentally observed TADs, comprising heterochromatic and active domains, while the former only captures the presence of PcG TADs~\cite{Sexton2012} (shown in the middle column). }
	\label{fullHiC}
\end{figure*}

\subsection{Including binding patterns of other proteins predicts location of remaining TADs}

The results discussed above pertain to the case in which only PRC1 interacted with the chromatin fiber. We refer to this case as the ``polycomb-only'' model from now on. There are other proteins which act as architectural players in shaping the 3D organization of chromatin. Heterochromatin associated protein %(\emph{e.g.}, SIR in yeast and %
HP1, %in higher eukaryotes) 
transcription factors and insulator proteins associated with enhancer-promoter loops are common examples. We therefore tackled the problem of understanding how these additional factors change the chromosome folding, which has been thus far driven by the PcG proteins alone. To this end, we introduced two other sets of binding and bridging factors, one representing a heterochromatin-binding factor, and another one representing a euchromatin-binding factor. We chose these generic binders for these epigenetic states because mechanistic aspects driving folding of these chromatin states are not as clearly understood as in the case of PcG proteins, which have been clearly shown to the compact and regulate the topology of chromatin~\cite{Francis}.

We marked chromatin beads according to their affinity for the three binders in our simulations: (i) PRC1, (ii) heterochromatin-binding factor and (iii) euchromatin-binding factor. While the affinity between chromatin and PRC1 was determined through ChIP-seq data as previously discussed, we used chromatin state data from Refs.~\cite{Filion2010,Karchenko2011} to determine the interactions with the other two factors, as follows. First, we used chromatin states from Ref.~\cite{Filion2010}, and stipulated that heterochromatin factors only bind (with weak affinity, $\epsilon_3$ = 2 $k_BT$ ) to ``black'' chromatin state, which labels generic heterochromatin (accounting for almost half of the \emph{Drosophila} genome). The heterochromatin factors may therefore model proteins like histone protein H1, AT-Hook protein D1, SUUR, and Su(Hw), all proteins found in black chromatin and associated with architectural roles or the establishment of higher order chromatin structure~\cite{Filion2010}. Second, we used the chromatin state model from Ref.~\cite{Karchenko2011} to identify enhancers (state 1 in Ref.~\cite{Karchenko2011}) and promoters (state 3), as chromatin regions where euchromatin factors bind strongly (with high affinity, $\epsilon_4$ = 9 $k_BT$ ). Such euchromatin factors may represent clusters of transcription factors, bridging factors like CTCF and polymerases which can loop the chromatin of active genes. We also allowed binding of euchromatin factors to transcribed regions (states 2, 4 in Ref.~\cite{Karchenko2011}) with weak affinity, $\epsilon_5$ = 3 $k_BT$. 
%[While we could have used the ``euchromatin'' state in Ref.~\cite{Filion2010} to identify regions where euchromatin factors bind, this characterisation would not single out promoters and enhancers among which the strongest bridging interactions should form: this is why we instead resorted to the more sophisticated characterisation in Ref.~\cite{Karchenko2011} for euchromatin.] 
We refer to the model with three sets of binders as the ``full-model''. For simplicity, we present in this case mainly results obtained with a mitotic-like initial chromosomal state (see preparation details in SI and Suppl. Movies M4-M6). This structure is qualitatively similar to the stretched state used for the polycomb-only model, but has additionally an internal looped structure matching that found in some models for mitotic chromosome organization~\cite{Rosa2008} (see SI).
 
We find that incorporation of additional bridges in the full model leads to formation of further TADs, while PcG driven TADs remain largely unaltered. This suggests that, to a first approximation, the bridges act mostly independently of one another. Fig.~\ref{fullHiC} shows a comparison between the predicted and experimental contact maps across different regions of chr3R. At 1.5--2.35 Mbp a big TAD flanked by smaller TADs appear in the full model. The polycomb-only simulations (from any initial condition) show instead a weak domain encompassing these, but with no well defined boundary. As shown by the comparison, these TADs clearly correlate with TADs from Hi-C data, and fall in the ``null'' class~\cite{Sexton2012}. Next at 12--13.4 Mbp, additional TADs flanking the central PcG TAD (highlighted by the light blue box) come up in the full model; these are absent in the polycomb-only model, and again they correlate well with contacts map from Hi-C. Interestingly, on the left hand side, one can further see that TAD splits into two subdomains, as in experiments. Towards the end of chr3R there is a region, located at 23--24.4 Mbp, which remains completely structureless in the polycomb-only simulations (top row), but instead folds into regular TADs in the full model, yet again matching well with the Hi-C data.

\begin{figure*}[t]
	\includegraphics[width=1\textwidth]{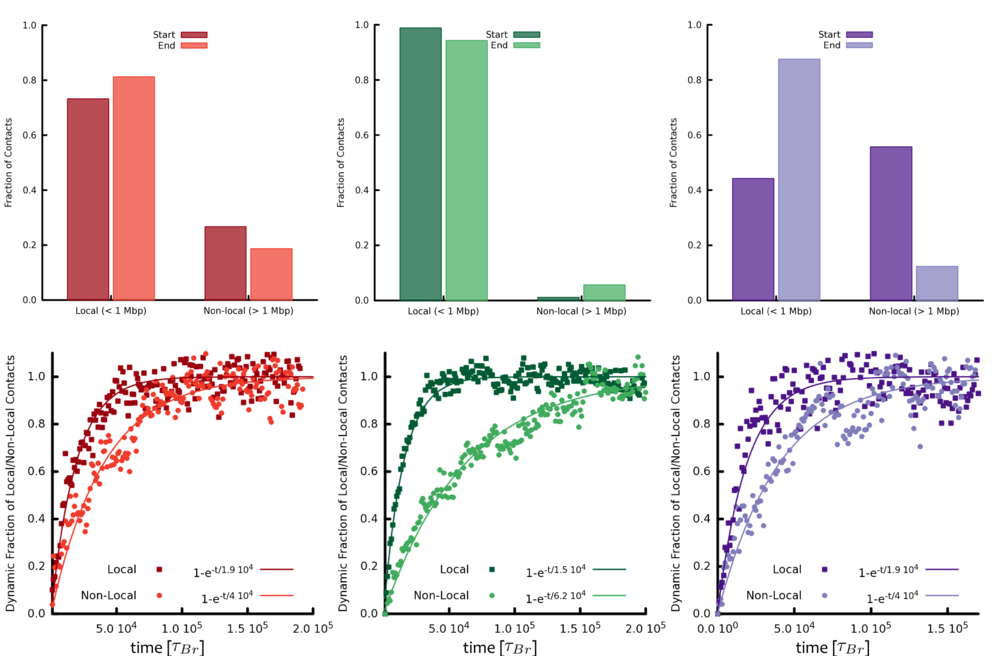}  %{../figures_3/fig1_plusFISH.eps}
	\caption{\textbf{Two-step folding of chr3R.} This figure shows the dynamic formation of ``local'' ($<1$ Mbp) and ``non-local'' ($>1$ Mbp) contacts in simulations of ``polycomb-only'' model. The top row shows the initial and final distribution of the contacts for the random walk (red, left), stretched (green, middle) and fractal globule (purple, right) initial configurations. The bottom row shows the time evolution of the number of local and non-local contacts from the initial state to the final one for the same initial chromosomal state. In all cases, there are two different time-scales, one associated with the fast formation of local contacts and another one associated with the slow formation of non-local contacts. This is strong evidence for a two-step mechanism for the folding of chromosomes exiting any arbitrary initial state. Here, $10^5$ $\tau_{Br}$ can be mapped to 1-10 minutes of real time (see SI).
	}\label{twostep}
\end{figure*}

To further validate the predicted contact map, 4C simulations using a bait at 12.77 Mbp on chr3R were performed (SI Fig.~S3). An asymmetric distribution of contacts was observed around the bait, which is expected given that the bait lies close to the boundary of the TAD (see Fig.~\ref{fullHiC} and SI Fig.~S3). This trend is well reproduced by both polycomb-only and full model simulations. Comparing with experimental 4C contact map, detailed analysis of the results shows another intriguing trend. The polycomb-only model shows four large peaks to the left of the bait which correlate with the 4C data, but are much larger than in experiments. Similarly, also the forward contacts predicted by this model are significantly more than those observed by 4C-seq. On the other hand, the number of contacts made by the bait decline in the case of the full model and the agreement with the experimentally observed 4C curve improves (SI Fig.~S3). Therefore, although at first approximation we can say that each of the bridges act separately to form separate TADs, the non-PcG bridges can screen PcG-mediated interaction, \emph{e.g.}, by sequestering weak binding sites for PRC1. Overall, both models give a good prediction for the position of troughs and peaks in the 4C experiment.
%(the most visible exception to this is marked by the red arrow in SI Fig.~S3).

From the above-presented simulations, it appears that 1D information obtained from ChIP-seq data can successfully be used to predict the chromatin contacts. Although this procedure can lead to a static picture of interphase chromatin organization, it is also crucial to understand the {\it dynamics} of formation of contacts and folding. In particular, how the interphase organization is attained by a chromosome upon exiting mitosis is still not clear. Understanding the relaxation of chromosomes from mitotic to interphase states may yield insight into how chromatin organization is inherited across cell generations and enabling cells to remember their identity. Our large scale simulations allow us to probe experimentally relevant time-scales and in light of this we proceed to analyse the chromosomal folding dynamics and study its features.

\begin{figure}[t]
	\includegraphics[width=0.5\textwidth]{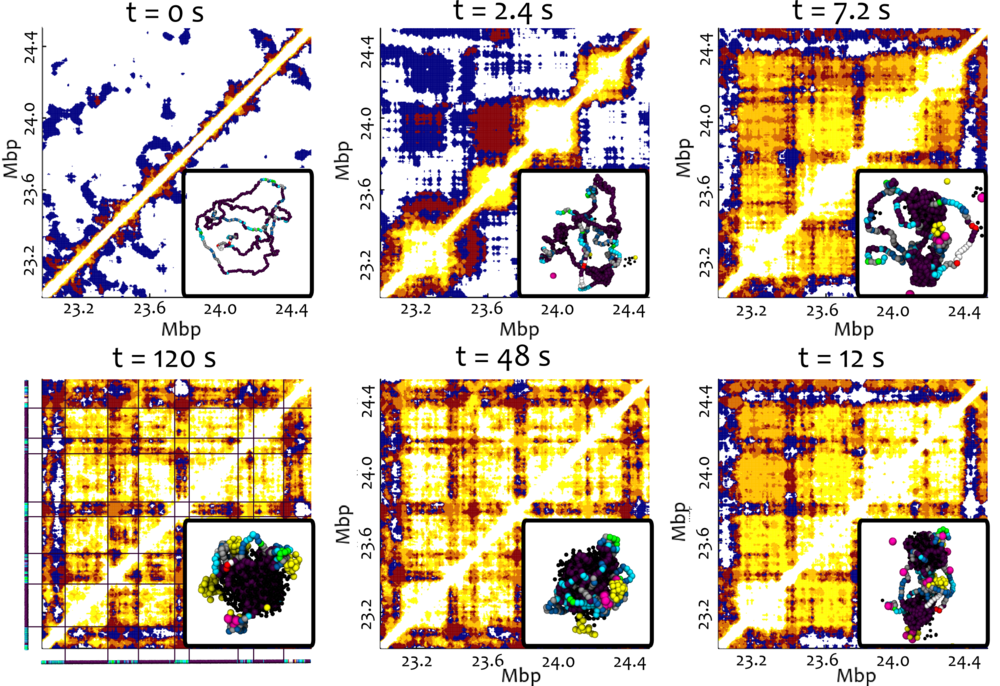}
	\caption{\textbf{The folding dynamics is arrested and displays long-lived TADs matching the experimental ones.} The time evolution of domains between 23 and 24.5 Mbp is shown. The contact maps are made by averaging over 6 independent simulations starting from a RW configuration and simulated using the ``full'' model. The snapshots represent configurations taken from one of the simulations. This figure clearly shows that local interactions and intra-TAD arrangements occur on faster time-scales than inter-TAD ones. It also suggests that the folding structure achieved at short time ($\sim 12$ seconds) is very long lived (virtually unchanged from 12 seconds on to 120 seconds) and well captures the epigenetic states along the polymer (see bottom left figure).}
	\label{fig:SI_Cmap_timelapse}
\end{figure}

\subsection{Chromatin folding is a two-step process}

In order to understand how chromosomes attain locus-specific interphase organization from a locus-independent mitotic state, we studied the relaxation dynamics of chr3R {\it in silico}. This process would be very difficult to probe experimentally as it would entail, for instance, an extensive Hi-C study of temporally synchronised cells. In light of this, simulations provide an ideal tool to gain some insight into this complex process.

Following the evolution of contacts from three distinct initial configurations using the polycomb-only model, different kinetics is observed for the formation of local and non-local contacts. For all three initial conditions, Fig.~\ref{twostep} shows that local contacts are established faster than non-local ones, and the growth rates also take similar values across different initial conditions. This suggests that the starting configurations chosen to model the initial state of chr3R do not affect its folding kinetics. Contacts are classified  as ``local'' when occurring between beads $<$ 1 Mbp apart along the genome, mainly due to intra-TAD interactions, while non-local contacts correspond to $>$ 1 Mbp interactions, mainly due to inter-TAD contacts. From these data it therefore appears that the folding of chr3R takes place in at least two stages. First, local TADs form, and then consolidation of the overall shape takes place by establishment of inter-TAD contacts. A further inspection of the time evolution of the contact maps (Fig.~\ref{fig:SI_Cmap_timelapse}) also suggests that local TADs are formed earlier than the establishment of non-local contacts. This can be readily seen in Fig.~\ref{fig:SI_Cmap_timelapse}, where we show the evolution of a selected region of chr3R simulated using the ``full model'' and starting from a random walk configuration. This case displays a quick local folding (within $\sim$ 5 seconds) followed by slower non-local arrangement (up to $\sim$ 12 seconds) which does not show further folding and suggests the establishment of a long-lived configuration (up to 2 minutes). Similar contact maps for chromosomes starting from mitotic-like configurations are reported in the SI (Figs.~S6-S7 and Suppl. Movies M4-M5).

\begin{figure}[t]
	\includegraphics[width=0.5\textwidth]{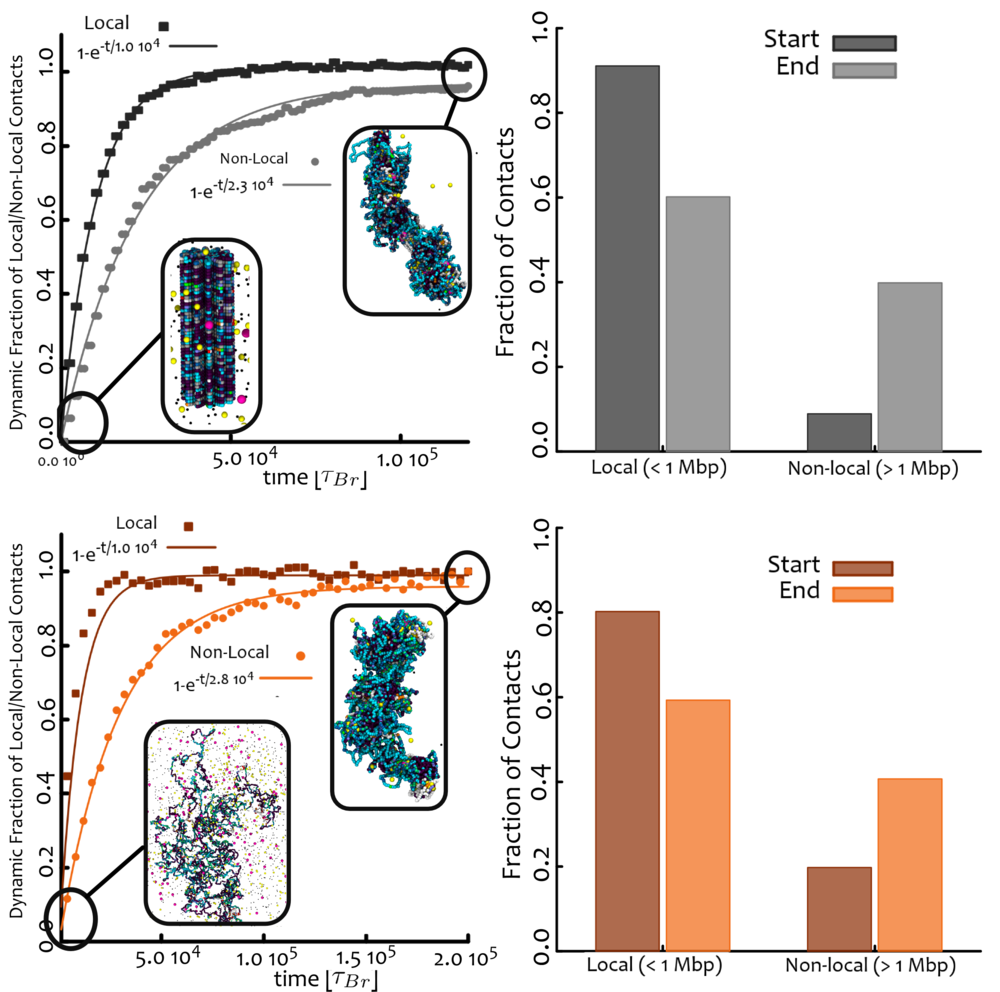}
	\caption{\textbf{The two-step folding process is also observed in the full model. } 
		\textbf{(left)} Kinetic of the formation of local and non-local contacts (defined and normalised as in Fig.~\ref{twostep}) for the full model, where chr3R is initialised in a mitotic cylinder configuration (top row) (see SI, Ref.~\cite{Rosa2008} and snapshot in inset) and as random walk (bottom row). Time is measured in simulation units (or Brownian times $\tau_{Br}$, see SI for mapping to physical units) and $10^5$ $\tau_{Br}$ can be mapped to 1-10 minutes of real time. \textbf{(right)} Histograms showing the non-local and local fraction of contacts in the simulated chromosome conformation at the start and end of the simulations shown for the two cases. The qualitative and quantitative similarity of the final configurations, distribution of contacts and folding dynamics is remarkable given so different initial configurations (compare snapshots in inset) and strongly suggest a very robust mechanism of genome organization.} 
	\label{fig:fulltwostep}
\end{figure}

\begin{figure*}[t]
	\includegraphics[width=0.90\textwidth]{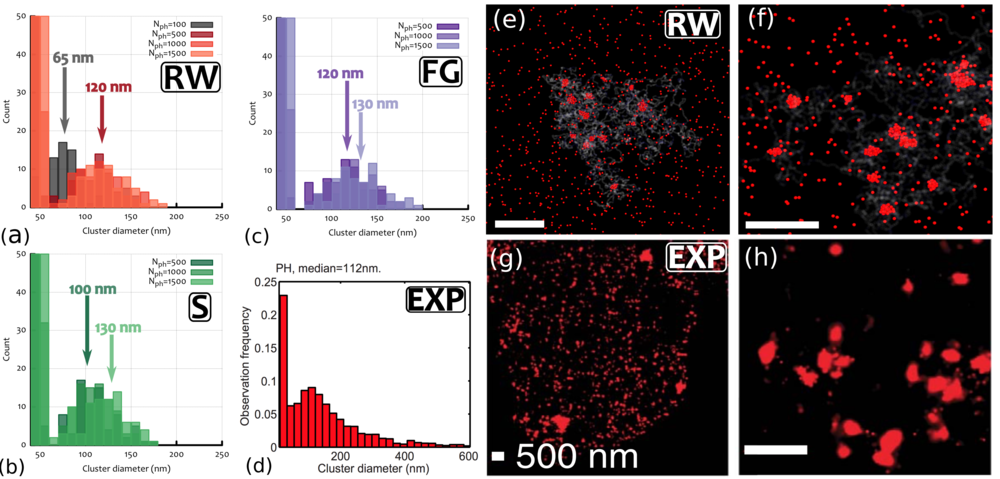}  %{../figures_3/fig1_plusFISH.eps}
	\caption{\textbf{Formation of PRC1 clusters in chr3R in computer simulations and STORM experiments.} \textbf{(a)}-\textbf{(c)} The final steady state distribution of cluster diameters in the three cases considered in this work and for different number of proteins $N_{ph}$ are shown. The three cases correspond to how we initialise the chromatin: as a random walk (RW), a stretched random walk (S) or a fractal globule (FG) (see SI). \textbf{(d)} The cluster diameter distribution measured from STORM experiments~\cite{Wani2016} shows a remarkable agreement with the simulations. \textbf{(e)}-\textbf{(f)} Snapshots from simulations showing the clustering of polycomb proteins (red). \textbf{(g)}-\textbf{(h)} Images form STORM experiment~\cite{Wani2016}. The scale bar is 500 nm in all images.   
	}
	\label{fig:Panel_clustering}
\end{figure*}

Using the full-model, we simulated folding of chr3R from a mitotic like conformation, which is one of the major chromatin reorganization events {\it in vivo}. Fig.~\ref{fig:fulltwostep} shows that also with this more complete model, and more realistic initial conditions, the conclusions on the kinetics of chromosome folding reached on the basis of the simpler polycomb only model remain unaltered. The chromosome appears to fold in a two-step fashion, with local contacts being established significantly earlier than non-local ones. Time evolution of contact map from mitotic-like state shows that non-local contacts continue to form after the establishment of local TADs (SI Fig.~S6-S7, Suppl. Movies M4-M6). The fraction of local contacts decreases while fraction of non-local contacts increases upon relaxation from the mitotic-like state. This appears to be in agreement with experimental observations~\cite{Naumova2013}, where application of Hi-C showed that mitotic chromosomes are depleted of long-range contacts, TADs, but enriched in very short-range (local) contacts. Here also the nature and genomic coordinates of contacts appears different between initial and final conformations. On the other hand, the dynamics of folding is again very robust against different initial conditions. In Fig.~\ref{fig:fulltwostep} we also show the folding dynamics of chr3R starting from a random walk initial configuration (see Suppl. Movie M7). This is both qualitatively and quantitatively strikingly similar to the dynamics of and final states reached by the mitotic-like starting chromosome (compare final snapshots in insets of Fig.~\ref{fig:fulltwostep} and the time-scales in the two cases). These results strongly suggest a broad robustness of the folding dynamics which might give a clue on the reproducibility and stability of TADs {\it in vivo}. 

Even though the kinetics are qualitatively similar for different initial configurations and the two models considered (compare Figs.~\ref{twostep} and \ref{fig:fulltwostep}), the quantitative balance struck between local and non-local contacts over time varies for the three different initial conditions. For the RW case, an increase in local and a decrease in non-local contacts is observed, while for the stretched configuration a moderate decrease in local contacts is seen. Finally, for the FG case, the fraction of local contacts  increases, while that of non-local ones decreases over time. This dynamic quantitative evolution is accompanied by an even more significant qualitative dynamic change in the nature and genomic coordinates of contacts, which are very different in the initial and final states across different starting conditions (compare top and bottom rows in Fig.~\ref{fig:Domains}). 

Given that the extent of non-local contacts varies across initial conditions, one might expect that upon waiting a very long time all initial conditions would eventually reach the same steady state properties for the non-local contacts. Even if this were the case, it is likely that, in practice, this time-scale may not be easily reachable in reality: for instance, the equilibration (Rouse) time of a polymer the size of chr3R in isolation (and under infinite dilution) is $10^8$ time units, or 100-1000 minutes, and nuclear crowding is likely to further slow down the dynamics dramatically~\cite{Rosa2008}. It is therefore possible that the system may still evolve, but on much larger time-scales; alternatively, it may be effectively arrested in a metastable state dependent on the initial condition. We favour the latter interpretation on the basis of our simulations, as no detectable change in the pattern is observed even in selected, longer simulations of $10^6$ time units. Furthermore, the radius of gyration and average cluster size appear to have reached a steady state within our simulation time, implying there might not be further major conformational changes (SI Fig.~S4-S5). This is also strongly supported by Figs.~S6-S7, where we show the time evolution of selected contact maps.

%Having established that the {\it local} polycomb domain structure found in our simulations is robust and does not depend on the starting configuration, we now compare it with the highest resolution Hi-C data available for {\it Drosophila} (10 kbp resolution, Ref.~\cite{Sexton2012}; note that, at this resolution, the existing experimental data actually mainly only probe these local contacts). Fig. 3 shows a zoom of the contact map found in simulations and in experiments (the region under consideration is that between 12 and 14 Mbp in chr3R). 

\subsection{Protein clusters form concomitantly with chromosome folding}

In all of the above-presented simulations we showed that chromatin associated proteins bind to the genome at specific sites, and then interact with each other to bridge different regions of chromatin. This implies that the distribution of chromatin-associated proteins will change as chromatin will reorganize. We therefore now analyse the dynamics of PRC1 during chromosome folding {\it in silico}, as the subnuclear distribution and dynamics of this protein complex are relatively well studied experimentally~\cite{Brand2010,Wani2016}.

%Using the full-model, we followed the dynamics of PRC1 during folding of chr3R: we observed that PRC1 started bridging different segments of chr3R by self-associating and this resulted in the formation of PRC1 clusters. A biphasic kinetics was observed for self-association of PRC1, a fast phase followed by slow phase reaching to steady-state (SI Fig.~S4). Assuming a size of 20nm for PRC1, we observed that cluster size reached 120 nm which is the range of experimentally observed values.

Using both models and for all initial conditions, we remarkably find that the PRC1 proteins rapidly form clusters, which then grow more slowly and appear to reach a stable steady state within our simulated time window, after which no appreciable coarsening occurs (Fig.~\ref{fig:Panel_clustering} and SI Fig.~S5). Although we considered several different possible values for the concentration of PRC1, we found that all these, except the smallest one (100 proteins for the whole of chr3R) led to very similar cluster size distributions. Remarkably, by assuming a diameter of about 30 nm per complex, these distributions compare well with those observed by STORM (Fig.~\ref{fig:Panel_clustering}). Because the affinities of PRC1 are only determined qualitatively from ChIP-seq data~\cite{Wani2016} (see SI), we further varied the values of high and low affinity sites used in the simulations, and found that, while clustering is observed for all the cases we considered, the distribution only matches well that observed by STORM for values of binding for high affinity  sites in the range 8-10 $k_BT$ and for weak sites in the range 2-4 $k_BT$ (see SI Fig.~S2). These values are reasonable estimates for DNA-protein interactions, which need to be in a range to allow stable but not irreversible binding. This also appears to be in agreement with about four-fold difference in binding intensity between low and high intensity ChIP-seq peaks (Fig. S1) The clustering occurs through the following positive feedback mechanism: the formation of a small cluster of PRC1 proteins is followed by an increased local concentration of polycomb group binding sites; because of this enhanced concentration, more PRC1 proteins are recruited there leading to a further increase in binding site density. This is an example of a generic mechanism through which DNA-binding proteins with multiple binding sites along the chromatin cluster, and which has previously been dubbed ``bridging-induced attraction''~\cite{Brackley2013,Johnson2015}, because it only works with proteins which can bridge chromatin (or indeed any other polymer to which they bind). In this case we consider that binding is specific, and the positive feedback loop we have outlined further requires that a single chromatin bead can bind to more than one protein complex. Because a chromatin bead represents 3.1 kbp (see SI, Methods), this is a realistic assumption for PRC1, as many of the ChIP-seq peaks are smaller than bead size and each of them can likely bind multiple molecules. We note that the affinity range we considered leads to stable rather than transient foci. This is in line with the slow recovery times observed for polycomb bodies; also note that post-translational protein modification, not included here, may potentially render foci dynamics faster. 

We followed the dynamics of PRC1 during folding of chr3R also using the ``full-model'' and observed that PRC1 started bridging different segments of chr3R by self-associating and this resulted in the formation of PRC1 clusters. A biphasic kinetics was observed for self-association of PRC1, a fast phase followed by slow phase reaching to steady-state (SI Fig.~S5). Assuming a size of 30 nm for PRC1, and the formation of spherical cluster (see SI), we observed that cluster diameter reached 120 nm which is the range of experimentally observed values.

In both the polycomb-only and the full model, the biphasic clustering dynamics of PRC1 mirrors the two-step chromatin folding. In particular, the time-scales associated with the biphasic dynamics compare well quantitatively with the time-scale associated with the faster local folding and the slower non-local folding of the chromatin fibre. The growth of PRC1 clusters also follow approximate power laws, where the apparent exponent is significantly larger for the faster initial dynamics (Fig.~S8). The apparent exponents are always lower than 1/3 at late times, which would be the exponent expected of diffusive coarsening for clusters of self-attracting particles~\cite{ChaikinLubensky}: possibly, this discrepancy is due to the fact that the coarsening droplets are made up by DNA-bound, rather than freely diffusing, proteins.

\section{Discussion and conclusions}

\begin{figure}[h]
	\includegraphics[width=0.45\textwidth]{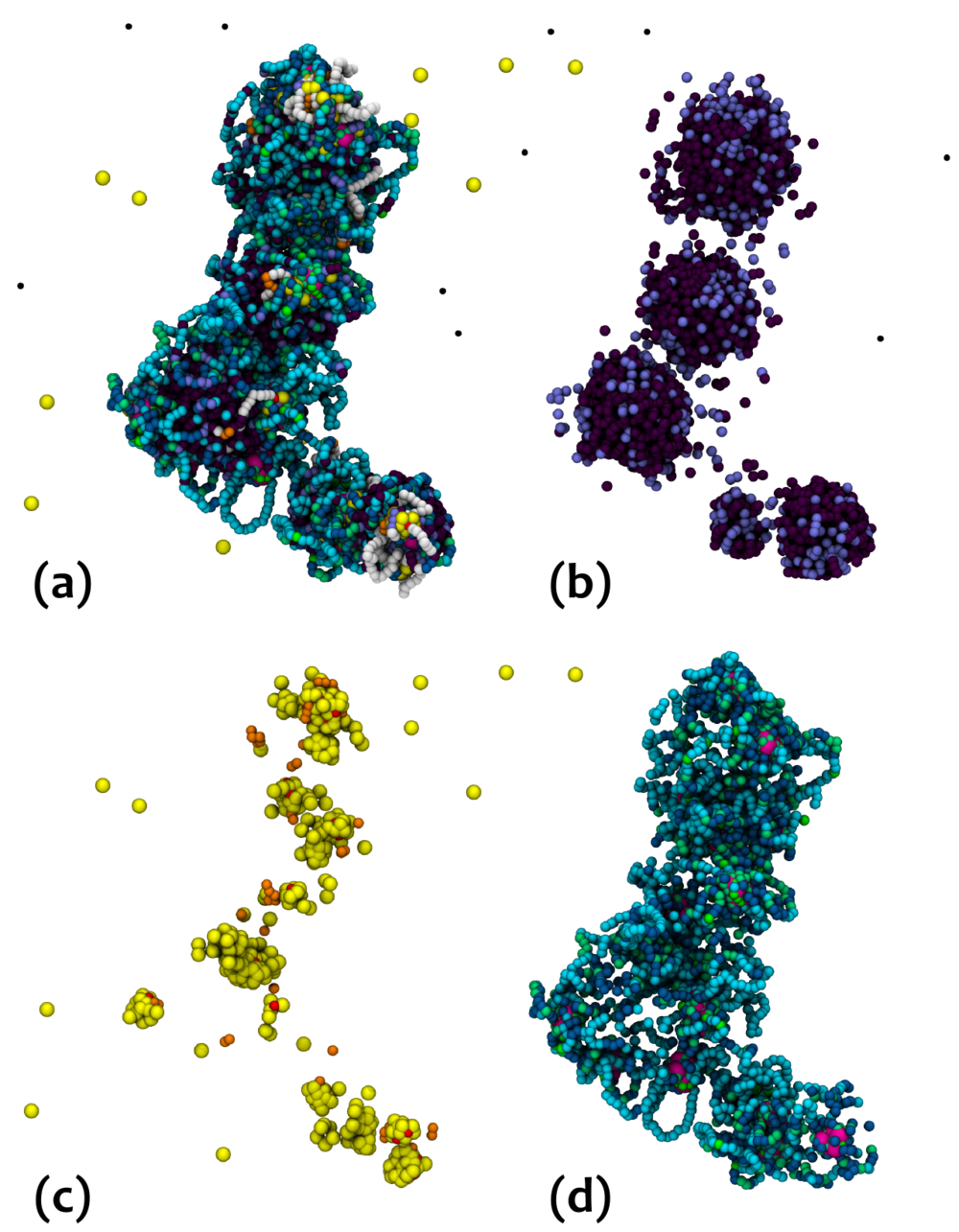}  
	\caption{\textbf{Snapshots from simulations of the ``full model''.}  \textbf{(a-d)} Show the final configuration of chr3R, as simulated via the ``full'' model (see text for details). Yellow, black and magenta spheres represent PRC1, heterochromatin-associating, and euchromatin-associating bridging proteins, respectively. White chromatin beads are non-binding; Chromatin beads coloured with hues of blue and hues of green denote weak and strong affinity to euchromatin-associating bridges; Red and orange chromatin beads have strong and weak binding affinity only for PRC1 proteins, respectively; beads coloured with hues of purple denote binding sites for heterochromatin-associating proteins. Beads representing eu- and hetero-chromatin can also have some (weak or strong) affinity to PRC1 proteins (resulting from the coarse graining procedure), and are therefore shown with different hues of colour. \textbf{(a)} All bead types are shown; \textbf{(b)} Shows heterochromatin-associating bridges (black) and their binding sites; \textbf{(c)} Shows PRC1 bridges (yellow) and their binding sites; \textbf{(d)} Represent active chromatin regions together with their associated protein bridges (magenta).}
	\label{fig:snapshots}
\end{figure}

In summary, we studied the folding dynamics of a \emph{Drosophila} chromosome (chr3R), covering realistic genomic and time scales, by means of Brownian dynamics simulations. These simulations are based on the simple assumption that chromosome structure is maintained by the action of ``bridges'': these are architectural proteins, or protein complexes, which can bind chromatin at many sites. We considered three types of chromatin-binding bridges: (i) PRC1, (ii) a generic heterochromatin-associating factor, and (iii) a generic euchromatin-associating factor, bridging enhancers and promoters. Prior to this full model, we also considered a simplified, polycomb-only, version where only the role of PRC1 was modeled. This allowed us to study in more detail the nature and assembly dynamics of polycomb-associated structures. Our modeling is informed by a series of sequencing data, on the basis of which we selected the affinity between a chromatin ``bead'' and a protein, these include ChIP-seq data (for PRC1), and chromatin state data (for heterochromatin- and euchromatin-associating factors). Our model can follow the evolution of both proteins and chromatin, and we compare the nuclear organization we find {\it in silico} with that found experimentally. An example of typical final configuration that we obtain from our simulations is represented in Fig.~\ref{fig:snapshots}(a), where we also show the typical 3D folding of heterochromatin (Fig.~\ref{fig:snapshots}(b)), PcG binding sites (Fig.~\ref{fig:snapshots}(c)) and euchromatin (Fig.~\ref{fig:snapshots}(d)).

An important outcome of our work is that the folding dynamics of eukaryotic chromosomes {\it in silico} proceeds in a two-step fashion. First, local domains (TADs) form fast (Figs.~\ref{twostep}-\ref{fig:fulltwostep} and SI Figs.~S6-S7) and reproducibly (independent of the initial chromosomal configuration, Fig.~\ref{fig:Domains}). Later on, non-local contacts form more slowly, and their structure markedly depends on the initial chromosomal state (Figs.~\ref{fig:Domains} and \ref{twostep}, \ref{fig:fulltwostep}). 
TADs have emerged as unifying feature of metazoan domains, but the mechanisms driving their folding are not clear. Our simulations suggest that interaction among chromatin  bound proteins can play a significant role in driving the folding of TADs. Observation of good correlation between predicted chromatin topology with that of experimentally observed topology demonstrates that 1D binding landscape of chromatin proteins can be helpful in predicting 3D organization.

This is an important difference with respect to the previous work on the folding of \emph{Drosophila} chromosome fragments (1 Mb in size)~\cite{Jost2014}, where significant dependence of the initial condition at the local (intradomain) scale was shown. The large-scale Brownian dynamics simulations reported in this work suggest instead a broader robustness with respect to the initial chromosomal conformation (see also Suppl. Movies M1-M3 and M6-M7). In this previous study~\cite{Jost2014} effective interactions between chromatin beads were only considered, whereas here we directly model the proteins which mediate such interactions. Therefore we are able to study the self-assembly of chromatin associated proteins like polycomb subnuclear clusters and we can follow directly the consequences of changing the nuclear concentration of such proteins. Furthermore, because we model the chromosome and associated proteins for biologically realistic time-scales, we can directly probe the chromosomal dynamics; on the other hand, in a scaled-down model kinetic features can only be indirectly accessed as the dynamics of the slowest dynamical processes, \emph{i.e.}, the formation of non-local interactions with genomic regions outside the simulation domain, are not included. 

In addition, while our results for the dynamics of long-range (non-local) contacts can be fitted with a time-scale of about several minutes (Fig.~\ref{twostep} and \ref{fig:fulltwostep}), the length of our simulation run-time allow us to reach a steady state, indicating that the chromosomes become trapped into very long-lived metastable states which can retain a memory of their conformational history. Our findings are therefore also consistent with the conjectured glassy dynamics of chromosomes~\cite{Rosa2008,Thirumalai2015}. More generally, our results suggest that the folding into local TADs is essentially an equilibrium process, driven by the interaction between chromatin and chromatin-associating bridging factors, while the same does not apply to the pattern of non-local contacts, and inter-domain interactions, which both depend more strongly on the initial chromosomal state and on stochasticity, leaving a signature of glassy dynamics (Fig.~\ref{fig:Domains}). 
%%DM added for point 1) explain history-dependence
\textcolor{black}{In other words, we find that inter-TAD interactions may depend on two main factors: (i) the chosen initial configuration, as one can notice from the full contact maps in Fig.~\ref{fig:Domains}, and (ii) the (stochastic) temporal sequence of long-range contacts which are established. The latter implies that different simulations may have different inter-TADs contacts as they follow independent trajectories in the phase space. This last observation is fully consistent with results obtained with single cell Hi-C, which displayed cell-to-cell variability due to intrinsic stochasticity~\cite{Nagano2013}. Nonetheless, we find that local interactions, or intra-TADs structures, are largely unaffected by such factors and this strongly supports the importance and stability of bridging-induced attraction in establishing local topological structures in interphase chromosomes.  }

Our simulations of the two-step relaxation of chr3R starting from a mitotic-like state can be used to suggest a possible model for the complex rearrangements that chromosomes must undergo during the cell cycle. In particular, the simulations predict a dynamics during which chromosomes first establish local contacts upon binding of chromatin proteins which are released from chromatin as cells enter into mitosis~\cite{Alberts}, and then, later on, non-local contacts form. 
The parent cell and daughter cells have the same chromatin organization, which in turn regulates gene expression, but during cell division huge reorganization of chromatin takes place. Understanding the mechanism and kinetic pathway by which interphase contacts are re-established after cells exit from mitosis can provide important information in order to address how memory for cellular identity is maintained across cell generations~\cite{Alberts}. Our simulations suggest that chromatin associated proteins play an important role in attaining the interphase organization after exiting mitosis. %and that this pathway is likely to depend both on the ability to retain epigenetic marks and on that of controlling the 3-dimensional configuration upon entry into interphase. 

In contrast to most of the simulations in the literature, as highlighted previously, we followed the dynamics of chromatin-binding proteins during chromosome folding. In our simulations, nuclear bodies made up of chromatin binding proteins arise naturally, through a generic positive feedback loop (also known as ``bridging induced attraction''~\cite{Brackley2013,Johnson2015,Brackley2015}). Notably, this feedback does not proceed indefinitely, and polycomb clusters only grow up to a steady state size in our simulations; the structural and kinetic features of these nuclear bodies also match very well with experimental observations. The size distribution of polycomb clusters in steady state {\it in silico} is close to that observed in \emph{Drosophila} S2 cells by STORM (Fig.~\ref{fig:Panel_clustering}). Furthermore, a polycomb cluster can self-assemble % from a pool of diffusing polycomb proteins 
in just 1-10 min in our simulations, which is similar to the assembly time found in vivo~\cite{Brand2010}. Strikingly, the self-assembly dynamics of the nuclear bodies also shows a two-step kinetics with power law growth, as well as arrested coarsening (see SI, Fig.~S5). 

Assumptions and predictions of our simulations appear reasonable as the final output correlates well with experimental observations. Overall good agreement is observed between the predicted and experimental contact maps (Fig.~\ref{fig:Domains} and SI Fig.~S2), which is even more enhanced when the full model is implemented (Fig.~\ref{fullHiC}). In addition, the comparison with the 4C-seq data (SI Fig.~S3) suggests that the three types of bridges which we consider share some of their binding sites, so that they compete with each other to bind to these, and, as a result, knocking out some of the proteins does not necessarily open up the associated chromatin domain. Therefore by incorporation of further one-dimensional information (from ChIP-seq data) better and more robust predictions can be made about three-dimensional chromatin organization.

In conclusion, the current work provides a framework within which to model in a realistic way structural and kinetic features of the 3D organization of chromatin and chromatin-binding proteins. We expect this approach to be generalized in order to predict how the organization is affected when the chromatin-protein binding affinity landscape changes, \emph{e.g.}, during development or in diseases, mainly based on the ChIP-seq data of the key bridging factors used here. In addition, in light of the observations reported here and of the size of the \emph{Drosophila} genome, we suggest that, with the aid of supercomputers, it is now theoretically feasible to follow, \emph{in silico}, the folding dynamics of entire eukaryotic genomes on realistic time-scale. This represent a crucial milestone towards achieving a more comprehensive understanding of the key mechanism leading the 3D organization of the eukaryotic nucleus.

\subsection{Simulations Details}
The entire chromosome 3R of \emph{Drosophila melanogaster} was simulated as a finitely-extensible and semi-flexible~\cite{Kremer1990} bead-spring co-polymer chain at a resolution of $30$ nm or $3.1$ kbp. The different type of beads composing the polymer represent the specific chromatin states as obtained from ChIP-seq tracks of \emph{Drosophila} S2 type cells~\cite{Sexton2012}. The persistence length of the chain is set to $l_p = 3\sigma \simeq 10$ kbp.
The dynamics of the monomers is evolved by means of a Brownian Dynamics (BD) scheme. The friction imposed on each monomer and the stochastic term added by the Brownian Dynamics scheme take into account the presence of a surrounding thermal bath whose temperature is set to $T=300$ K (see SI for further details). The Langevin equation is then integrated using the LAMMPS package in the canonical (NVT) ensemble. The systems were simulated in dilute regime (volume fraction $\rho=N\pi\sigma^3/6Vol \simeq 0.05\%$) and confined in a box to avoid self-interactions through the periodic boundaries. The binding proteins are modelled as spheres with diameter $\sigma = 30$ nm; these interact sterically with other proteins and the chromatin via a cut-and-shift LJ potential (WCA) allowing for an attraction region between specific chromatin states and corresponding binding proteins. The affinity of the binding proteins to specific chromatin \emph{loci} is varied to probe the robustness of the results (see SI for details). The initial configurations we considered are the most addressed in the literature and range from the self-avoiding random walk to a fractal globule state to model a random interphase conformation and from an elongated (persistent) random walk to a collapsed stack of rosettes to model possible post-mitotic situations (see SI for further details).

\section{Acknowledgements}
DMa and DMi acknowledge support from the ERC (Consolidator Grant THREEDCELLPHYSICS, Ref. 648050). AHW is a Ramanujan fellow, awarded by Department of Science and Technology, India.

\newpage
\appendix*

\setcounter{figure}{0}
\makeatletter 
\renewcommand{\thefigure}{S\@arabic\c@figure}
\makeatother

\section{Supplementary Material}
\section{Simulation Details}
The finitely-extensible worm-like chain is modelled via the Kremer-Grest model~\cite{Kremer1990}
as follows:
Let  {$\bm{r}_i$} and  {$\bm{d_{i,j}} \equiv \bm{r}_j - \bm{r}_{i}$}  be respectively the position of the center of the  {$i$}-th bead and the vector of length  {$d_{i,j}$} between beads  {$i$} and  {$j$}, the connectivity of the chain is treated within the finitely extensible non-linear elastic model with potential energy, 
\begin{equation}
U_{FENE}(i,i+1) = -\dfrac{k}{2} R_0^2 \ln \left[ 1 - \left( \dfrac{d_{i,i+1}}{R_0}\right)^2\right]  \notag
\end{equation}
for   {$d_{i,i+1} < R_0$} and  {$U_{FENE}(i,i+1) = \infty$}, otherwise; here we chose  {$R_0 = 1.6$ $\sigma$} and  {$k=30$}  {$\epsilon/\sigma^2$}.
The bending rigidity of the chain is captured with a standard Kratky-Porod potential,
\begin{equation}
U_b(i,i+1,i+2) = \dfrac{k_BT l_p}{\sigma}\left[ 1 - \dfrac{\bm{d}_{i,i+1} \cdot \bm{d}_{i+1,i+2}}{d_{i,i+1}d_{i+1,i+2}} \right],\notag
\end{equation}
where we set the persistence length $l_p = 3 \sigma$ in order to reproduce the observed stiffness of chromatin, \emph{i.e.} $l_p \simeq 100$ nm.\\

The steric interaction between monomer  {$a$} and monomer  {$b$} (of sizes  {$\sigma_a$} and  {$\sigma_b$}, respectively), representing either a chromatin region, or bridge proteins is taken into account via a truncated and shifted Lennard-Jones (or WCA) potential  
\begin{equation}
U_{LJ}(i,j) = 4 \epsilon_{ab} \left[ \left(\dfrac{\sigma_c}{d_{i,j}}\right)^{12} - \left(\dfrac{\sigma_c}{d_{i,j}}\right)^6 - \left(\dfrac{\sigma_c}{r_c}\right)^{12} + \left(\dfrac{\sigma_c}{r_c}\right)^6 \right] \text{ for } d_{i,j}< r_c  \notag
\end{equation} 
and 0 otherwise and where  {$\sigma_c=(\sigma_a+\sigma_b)/2$} is the minimum distance between beads $a$ and $b$ and {$r_c$} is the chosen cut off. This parameter is set to {$r_c=2^{1/6}\sigma$} in order to model purely repulsive interactions and instead set to {$r_c=1.8\sigma$} to include attractive interactions. In both cases, the potential is shifted to zero at the cut-off in order to have a smooth curve and avoid singularities in the forces.  
Purely repulsive interactions, such as those between proteins, have $\epsilon_{ab}=1$, while attractive interactions such as those between PRC1 proteins and Polycomb binding sites have been varied to study the robustness of the results. The range of parameters are reported in Table~S1. \\ %\ref{tab:values}.
\begin{table}[t]
\begin{tabular}{c|c|c}
Chromatin state & Binding Protein  & $\epsilon_{ab}$ [$k_BT$] \\
\vspace*{-0.2 cm} & & \\
\hline
\hline 
\vspace*{-0.05 cm} & & \\
PRC1 strong~\cite{Wani2016} & PRC1 & 6-10 ($\epsilon_1$)\\ 
PRC1 weak~\cite{Wani2016} & PRC1 & 2-5 ($\epsilon_2$)\\ 
Heterochromatin~\cite{Filion2010} & HF & 2 ($\epsilon_3$)\\ 
Euchromatin strong (States 1 3)~\cite{Karchenko2011} & TF & 9 ($\epsilon_4$) \\ 
Euchromatin weak (States 2 4)~\cite{Karchenko2011} & TF & 3 ($\epsilon_5$)\\ 
\end{tabular}
\caption{Table of the parameters {$\epsilon_{ab}$} used to model the attractive interaction between DNA and specific binding proteins. The polycomb binding sites are taken from Ref.~\cite{Wani2016}. The eu- and hetero-chromatin states are obtained from the works Ref.~\cite{Filion2010,Karchenko2011} (see main text). The strong and weak affinities of the PRC1 binding sites are varied to explore a range of cases. The strongly and weakly binding euchromatin represent strong enhancer-promoter interactions and weaker interactions in the other cases (such as generally active euchromatin), respectively. These can be bound to generic transcription factors (TF). Heterchromatin is instead bound by generic heterochromatin factors (HF).}
\label{tab:values}
\end{table}\\
\begin{figure*}[h]
	\includegraphics[width=0.75\textwidth]{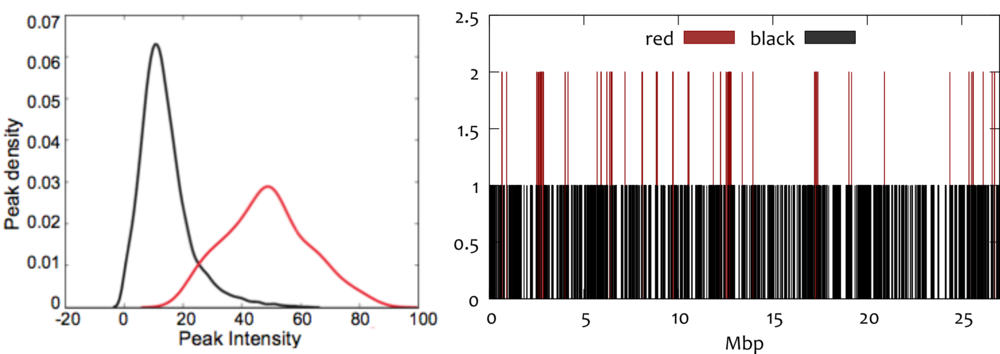}
	\caption{\textbf{(left)} Density of binding sites in \emph{Drosophila} genome as a function of their intensities. One can notice two distinct types of binding sites for a PcG protein, PH. These strong (red) and weak (black) binding sites were identified previously~\cite{Wani2016} and sequencing data was deposited in NCBI’s Gene Expression Omnibus and accessible under accession number GSE60686. \textbf{(right)} ChIP-seq (binding) profile for strongly (red) and weakly (black) binding sites on \emph{Drosophila} chr3R. The y-axis is arbitrary in that the strong (red) binding sites have higher values than the black ones only for visualisation purposes. The affinity of the binding sites is tuned to study a range of cases (see text), while the position of the binding sites reported here is carefully mapped to the coarse-grained polymer model.}
	\label{fig:chipseq}
\end{figure*}

In order to avoid topologically frustrated configurations we allow the beads forming the chain the pass through themselves with a small probability. This is regulated by a soft potential of the form
\begin{equation}
U_{\rm soft}(i,j) = A\left[ 1 + \cos{\left( \dfrac{\pi d_{i,j}}{r_c} \right)} \right] \theta(r_c - d_{i,j})
\end{equation}
with $A=20 k_BT$ being the maximum height of the potential barrier, \emph{i.e.}, crossing probability $p = \exp{-20}$. 

Denoting by $U$ the total 
potential energy, the dynamic of the beads forming the polymer is described by the following Langevin equation:
\begin{equation}
m \ddot{\bm{r}}_i = - \xi \dot{\bm{r}}_i - {\nabla U} + \bm{\eta}
\label{langevin}
\end{equation}
where $\xi$ is the friction coefficient and $\bm{\eta}$ is the stochastic delta-correlated noise. The variance of each Cartesian component of the noise, $\sigma_{\eta}^2$ satisfies the usual fluctuation dissipation relationship $\sigma_{\eta}^2 = 2 \xi k_B T$.

As customary~\cite{Kremer1990}
we set $m/\xi = \tau_{LJ}=\tau_{Br}$, with the LJ time $\tau_{LJ} = \sigma \sqrt{m/\epsilon}$ and the Brownian time $\tau_{Br}=\sigma/D_b$ is chosen as simulation time step. Here, $D_b = k_BT/\xi$ is the diffusion coefficient of a bead of size $\sigma$. From the Stokes friction coefficient of spherical beads of diameters $\sigma$ we have: $\xi = 3 \pi \eta_{sol} \sigma$ where $\eta_{sol}$ is the solution viscosity.
One can map this to real-time units by using the viscosity of nucleoplasm, which ranges between $10-100$ cP, and by setting $T=300$ K and $\sigma=30$ nm, as previously defined. From this it follows that $\tau_{LJ} = \tau_{Br} = {3 \pi \eta_{sol} \sigma^3/\epsilon}\simeq 0.6-6$ ms. 
The numerical integration of Eq.~\eqref{langevin} is performed by using a standard velocity-Verlet algorithm with time step $\Delta t = 0.01 \tau_{Br}$ %\sim 0.4$ $ns$ 
and is implemented in the LAMMPS engine. We perform simulations up to $2$ $10^5$ $\tau_{Br}$ which correspond to 2-20 minutes in real time. 
The thermal energy  {$k_BT$} is set to  {$\epsilon$}, the energy scale of the WCA potential for the purely repulsive interactions. \\

\subsection{Modelling the Affinities between Proteins and Chromatin}
For the ``polycomb-only'' model we only consider protein PRC1 and PcG  binding sites; this means that the polymer is made up of beads that can be one of three types: (i) non interacting, (ii) strongly binding to PRC1 ($\epsilon_1$) or (iii) weakly binding to PRC1 ($\epsilon_2$). The polymer is then initialised in a box together with a random distribution of PRC1 proteins, here modelled as spheres of diameter $\sigma$ which can then freely interact to the polymer. In the full model the beads can be in one of the 12 possible states which are made by considering that a bead is a coarse-grained section of chromatin ($3.1$ kbp) and can therefore encompass more than one chromatin state. If the two states are not mutually exclusive (such as euchromatin and heterochromatin for which only the more abundant state is chosen as the marking type for that bead) then a bead can have have an affinity for both, PRC1 and HP1 or PRC1 and TF. For the first case, one can choose between the 3 combinations of affinities ($\epsilon_1$, $\epsilon_2$ or no affinity) for PRC1 and one for the heterochromatin $\epsilon_3$. For the second case, one has to combine three PRC1 affinities and two TF affinities ($\epsilon_4$,$\epsilon_5$)(see table T1). In total we therefore have 9 mixed types, one neutral and 2 PRC1-only affine. Alongside, three proteins identifying generic binding proteins (PRC1, HP1 and TF) complete the system.  

\subsection{Pre-Equilibration}
In order to avoid numerical blow-up and unwanted metastable states dominated by entangled (knotted) configurations, we model the interaction between beads forming the chromatin via a stiff soft potential, \emph{i.e.} 
\begin{equation}
U_{\rm pre-soft}(i,j;t) = A_{\rm pre}(t)\left[ 1 + \cos{\left( \dfrac{\pi d_{i,j}}{r_c} \right)} \right] \theta(r_c - d_{i,j}).
\label{eq:SoftPot}
\end{equation} 
with $A_{\rm pre}(t)$ a ramp function increasing from 0 to 20 $k_BT$ during the pre-equilibration time. This potential allows for frequent strand-crossing events at the beginning of the pre-equilibration and progressively forbids them in time. This aims to reproduce the presence of topological enzymes such as topoisomerases, which can un-entangle knotted chromatin conformations, especially close the mitotic state. The soft potential (with $A_{\rm pre}(t)=A=20 k_BT$) is also employed later on during the simulations to allow rare strand-crossings due to particularly entangled and frustrated configurations and mimics the presence (although at smaller concentrations) of Topo II.

\section{Initial Configurations}
\begin{figure*}
	\includegraphics[width=0.75\textwidth]{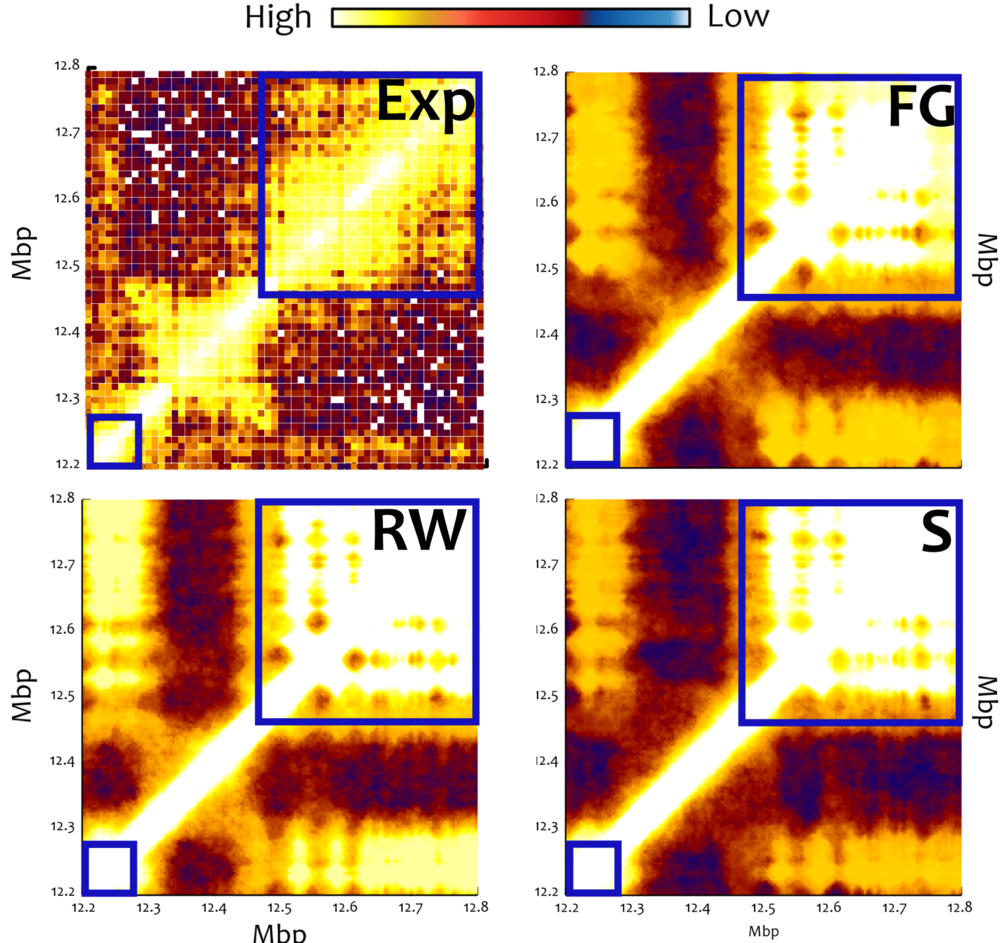}  %{../figures_3/fig1_plusFISH.eps}
	\caption{  These plots show the contact maps found experimentally \textbf{(Exp)}, and by simulations \textbf{(FG-RW-S)} for the region 12-13 Mbp. The same colorbar applies to all plots and is shown at the top. The simulated contact maps correspond to different initial conditions: fractal globule \textbf{(FG)}, random walk \textbf{(RW)}, stretched configuration \textbf{(S)}. As these results are reported form the ``polycomb-only'' model, it is important to notice that the PcG TADs are the ones highlighted by the blue squares, and therefore our model correctly captures those, and {\it not} the domain in the middle. The local folding is robust and essentially independent of the initial condition and also shows additional contacts between the PcG domains in agreement with the experiments.
	}
	\label{fig:Sexton_feature}
\end{figure*}

In this work we four different classes of initial conditions: (i) Self-avoiding Random Walk (RW), (ii) Stretched (S), (iii) Fractal globule (FG) and (iv) Mitotic (M). The RW configuration of a  {$N$}-beads long polymer is generated by standard means in which every bond connecting two beads is picked in a random direction, uncorrelated with respect to the previous bond. This is then turned into a self-avoiding polymer by performing a short run in which the beads are subject to the soft potential of eq.~\eqref{eq:SoftPot}. The stretched configuration is generated by biasing the RW algorithm in order to display a larger persistence length along the  {$\hat{z}$} direction and by performing the short soft repulsion run. The FG configuration is generated via a quick collapse of a RW configuration. This is done by extending the cut-off of the WCA potential describing the interaction between the polymer beads. In this way we induce a fast non-equilibrium collapse. A confining sphere is also used to drive the symmetric collapse of the polymer which would, otherwise, display segregated (dumbbell-like) sections. The mitotic conformation has been modelled by using the equations reported in Ref.~\cite{Rosa2008}. The following generalised helix describes a stack of rosettes of length $l_r = 200 \sigma \simeq 600$ kpb with 12 ``leaves'' each, meaning that each loop forming a rosette is around $50$ kbp long. The equations are
\begin{align*}
x(\phi) &= r_{chr} \left[ f+(1-f)\cos{\left(k\phi\right)}^2\cos{\phi}\right] \\
y(\phi) &= r_{chr} \left[ f+(1-f)\cos{\left(k\phi\right)}^2\sin{\phi}\right] \\
z(\phi) &= \dfrac{p\phi}{2\pi} 
\end{align*}
where $f  \times r_{chr}= 0.38 \times 19 \sigma \simeq 7.22\sigma \simeq 200$ nm is the radius of the mitotic cylinder, $p$ is the vertical step for each full turn and $k=6$ in order to get 12 leaves per rosette. This means that $9002$ beads, forming a polymer with contour length $L_c=9000\sigma \simeq 27.9$ Mbp, are laid into planes of $200$ beads each and with separation of $\sigma$ between planes, leading to a cylinder of height $h\simeq 1.3$ $\mu$m. \\
Interestingly, we observed that the local folding dynamics is largely independent on the chosen initial configuration. An example is reported in Fig.~\ref{fig:Sexton_feature} where we compare a PcG topologically associated domain folding on sub-megabase scales and consistently recovered for several choices of initial conditions and well matching with the experimentally observed structure.    
On the other hand, we observed that the non-local, long range, folding is instead affected by the choice of initial condition. This can be seen from the full contact maps reported in Fig.~1 of the main text.  

\section{Observables}
The most important observables that we keep track of are the contact maps, radius of gyration and clustering of PcG proteins. These quantities are reported in main text and SI, and if not otherwise specified, they are produced by averaging over 6 independent replicas. The contact maps are obtained using a cut-off of $3\sigma$ for 3D neighbouring beads. The radius of gyration computed as 
\begin{equation}
	R_g^2 = \dfrac{1}{2N^2}\sum_{i=1}^N\sum_{j=1}^{N} [\bm{r}_i - \bm{r}_j]^2.
\end{equation}
The clustering analysis is performed setting a cut-off distance $d_c$ of $2\sigma$, for which two beads are in the same cluster only if their distance is smaller than $d_c$. All other observables, can be obtained by further processing these quantities. 

\section{Robustness of interaction parameters}
The robustness of the results is probed by repeating the simulations for different values of $\epsilon_1$ and $\epsilon_2$. 
We performed the test only for the ``polycomb-only'' case. We observed that the cluster distribution is consistent with experiment (see Fig.~6 in main text), \emph{i.e.} it shows a peak at cluster diameters  at {$d_C\simeq 120$} nm, for a rather spread region of the interaction parameters {$\epsilon_1$} and  {$\epsilon_2$}. We explored the ranges  {$\epsilon_1 \sim 6-10$} and  {$\epsilon_2 \sim 2-5$}. In all the cases we could observe the formation of clusters with diameters ranging between  {$100-150$} nm. Nonetheless, the nature of the clusters changed when the difference between  {$\epsilon_1$} and  {$\epsilon_2$} was in the region of few $k_BT$s or $\epsilon_2$ was larger than {$4 k_BT$}. In these regimes the clusters could be seen to contain a larger fraction of weakly polymerising polycomb binding sites. At the same time, the contact probability of monomers $s$ segments apart would show a power law behaviour strongly deviating from $s^{-1}$. This is reported for one case where $\epsilon_1=10$ $k_BT$ and $\epsilon_2=5$ $k_BT$ in Fig.~\ref{fig:SI_Clusters}. 

%\section*{Sexton Thistle}
%To further quantitatively compare our model with experimental results, we report a close-up section of the chromosome 3R ranging between 12.2 and 12.7 Mbp in fig.~\ref{fig:Sexton_Thistle}.
%\begin{figure}[t]
%\centering
%\includegraphics[width=0.45\textwidth]{../Figs/SextonThistle_RW500_PcG_yellowHue.png}
%\caption{Close up section of the Chromosome 3R as reported in Sexton et al. (upper half) and as obtained \emph{in silico} in this work (bottom half).}
%\label{fig:Sexton_Thistle}
%\end{figure} 

\section{Simulated 4C experiments}
\begin{figure*}[t]
 	\includegraphics[width=0.95\textwidth]{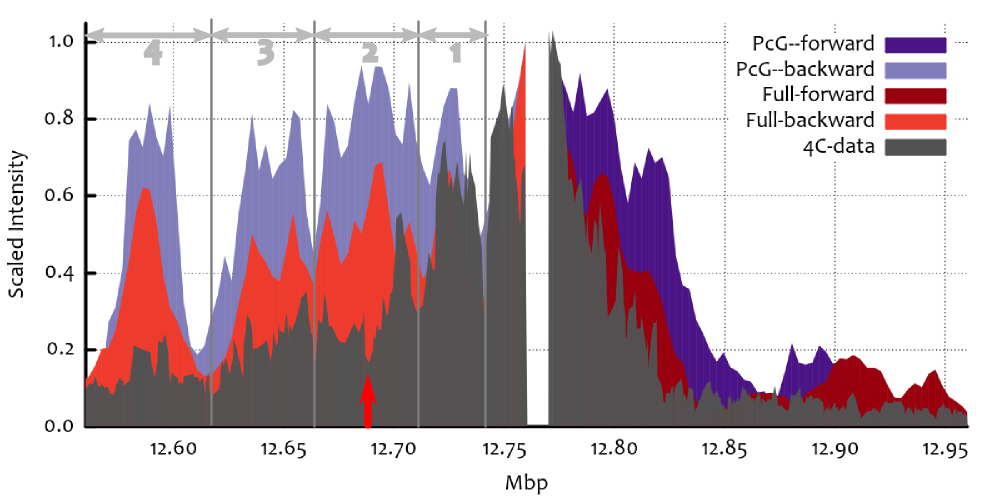}  
 	\caption{\textbf{Comparison between simulations and 4C-seq data. } The simulated 4C experiment is in qualitative agreement with the experimental profile (dark grey, from Ref.~\cite{Wani2016}). Both models, ``polycomb-only'' (hues of blue) and ``full'' (hues of red), qualitatively capture the asymmetric profile while the latter model also provides some clues on how the competition between binding proteins might play a role in determining the chromatin folding (see discussion in main text).    
 	}\label{4C}
 \end{figure*}

In order to compare our model with experiments we also perform a simulated 4C experiment. By placing a ``bait'' at 12.77 Mbp, we can record the contact profile along the rest of the chromosome (Fig.~\ref{4C}). Interestingly, the forward and backward profile are not symmetric with respect to the bait. In particular, the backward profile is generally higher than the forward one, indicating that the location of the bait is at the end of a TAD. In addition, we observe that both, ``PcG-only'' and ``full'', models capture the general behaviour of the profile, although the latter is closer to the experimentally probed intensity. This can be due to the fact that considering more proteins types and possible binding sites creates a competing effect that lowers and redistributes the chromosome contacts.   
 
\begin{figure*}[h]
	\centering
	\includegraphics[width=1\textwidth]{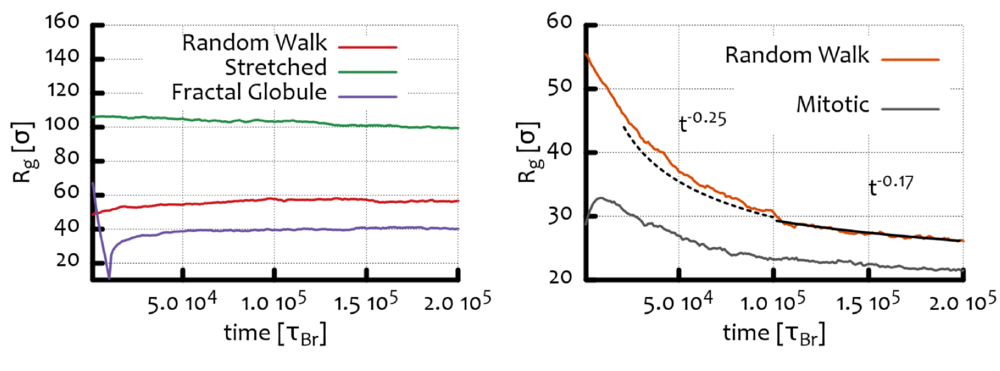}
	\caption{Evolution of the radius of gyration $R_g$ of the chromosome ch3R in the case of the ``polycomb-only'' (left) and and ``full'' (right) model.  The different initial configurations are indicated in the figure legends. The ``full'' model drives a stronger compaction, possibly due to the more numerous proteins considered. The fractal globule case (the dramatic decay indicates the artificial collapse of the polymer prior to its release) shows instead a swelling that contributes to losing many local contacts. }
	\label{fig:SI_coarsening}
\end{figure*} 

\section{Coarsening} 
\label{sec:coarsening}
\begin{figure*}[t]
	\centering
	\includegraphics[width=1\textwidth]{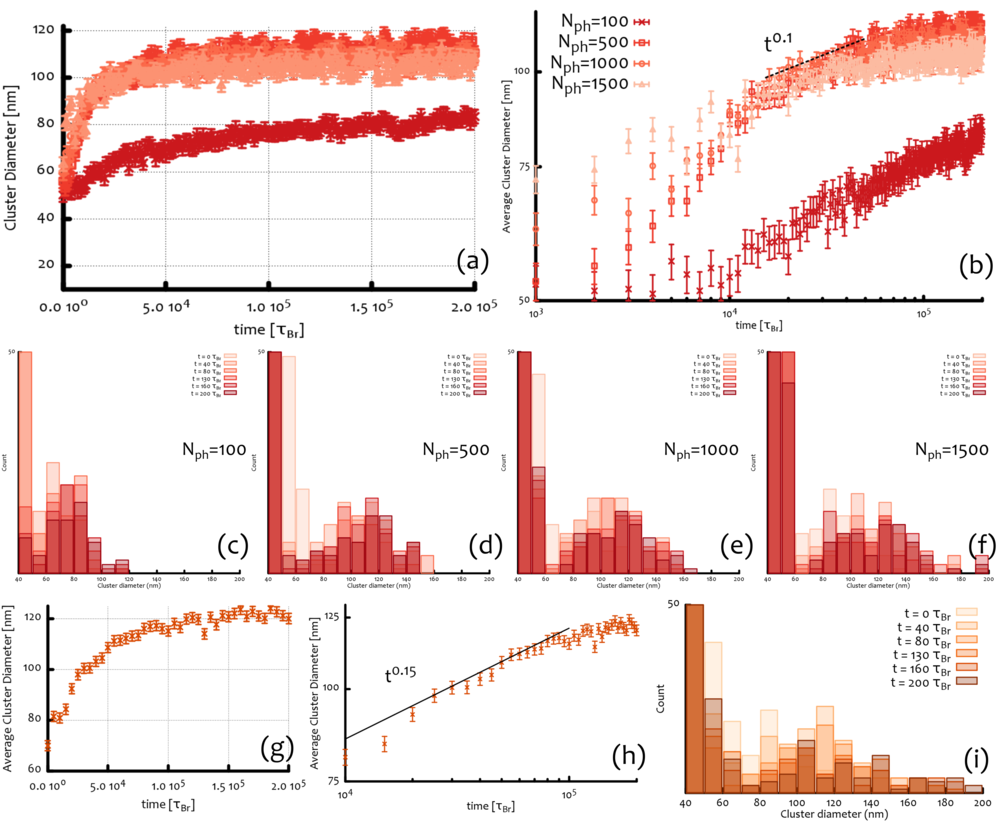}
	\caption{\textbf{(a-b)} Time evolution of the average cluster size. The coarsening dynamics of the binding proteins slows down towards the end of the simulation, indicating that the systems has reached a (quasi-)steady state. \textbf{(c-f)} Time evolution of the cluster size for the case of the ``polycomb-only'' model starting from a random walk configuration and with different number of PH proteins ($N_{ph}$). The cluster size count reaches a steady state whose average is around $120$ nm, remarkably close to the one experimentally observed (see main text). \textbf{(g-i)} Coarsening dynamics for the full model. The clusters of PRC1 display a severely slow coarsening towards the end of the simulation also in this case.  }
	\label{fig:SI_Cluster_coarsening}
\end{figure*}

\begin{figure*}[t]
	\centering
	\includegraphics[width=1\textwidth]{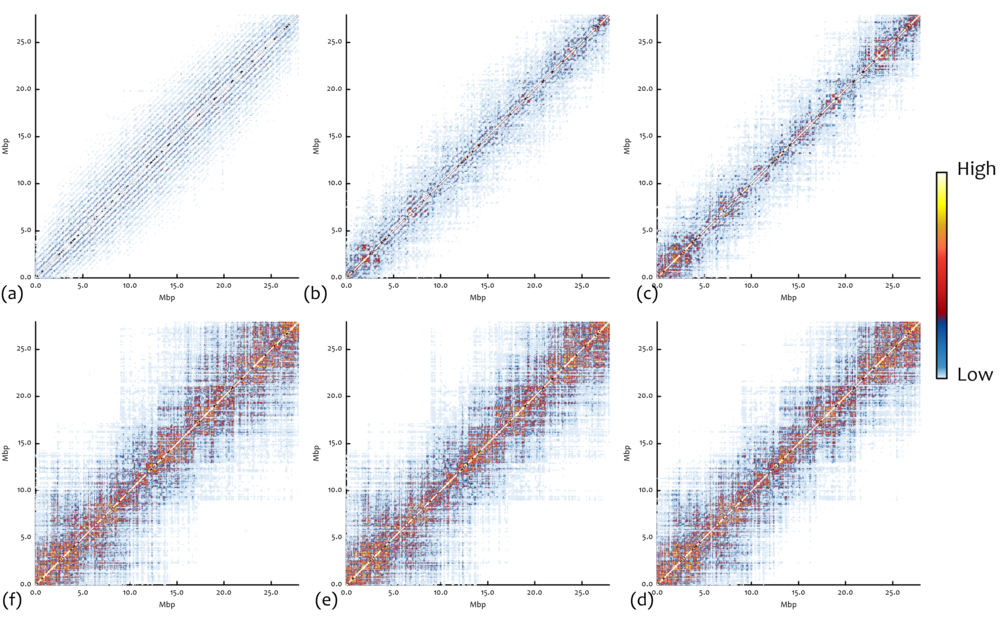}
	\caption{Time evolution of the whole contact map staring from a mitotic-like configuration. The snapshots are taken at $t=0$, $2.4$, $4.8$, $38.4$, $91.2$ and $120$ seconds (for viscosity $\eta= 10$ cP).  One can notice a very defined rosette structure in (a) that is lost to the advantage of the formation of TADs. See Movie M4.}
	\label{fig:SI_coarsening_Cmap_M}
\end{figure*}

\begin{figure*}[t]
	\centering
	\includegraphics[width=1\textwidth]{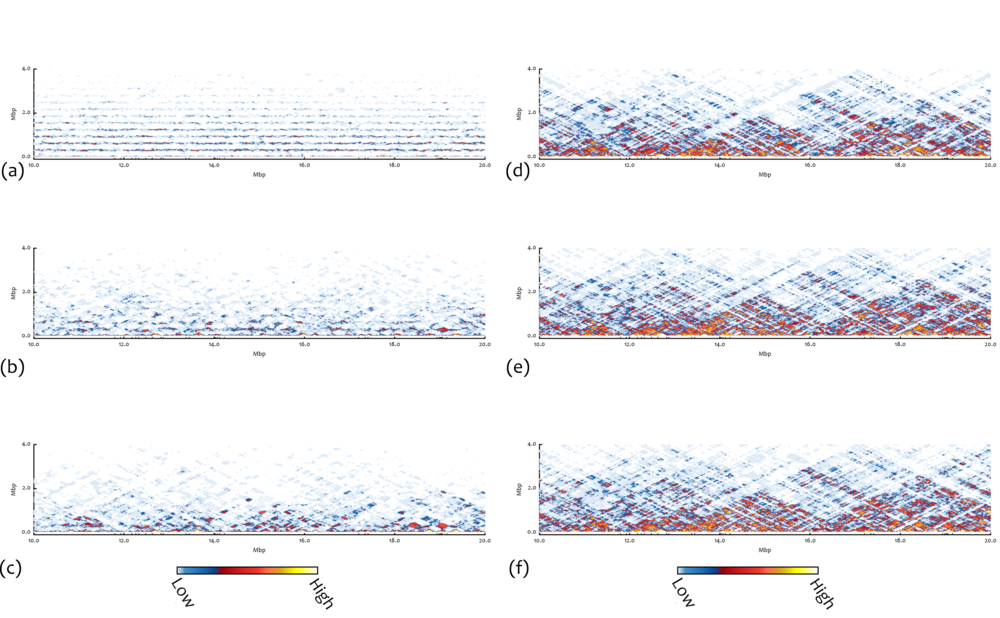}
	\caption{Time evolution of domains between 10 and 20 Mbp starting from a mitotic-like configuration. The snapshots are taken at $t=0$, $2.4$, $4.8$, $38.4$, $91.2$ and $120$ seconds (for viscosity $\eta= 10$ cP). One can notice a very defined rosette structure in (a) that is lost to the advantage of the formation of TADs. See Movie M5.}
	\label{fig:SI_coarsening_Dom_M}
\end{figure*}

\begin{figure*}[t]
	\centering
	\includegraphics[width=\textwidth]{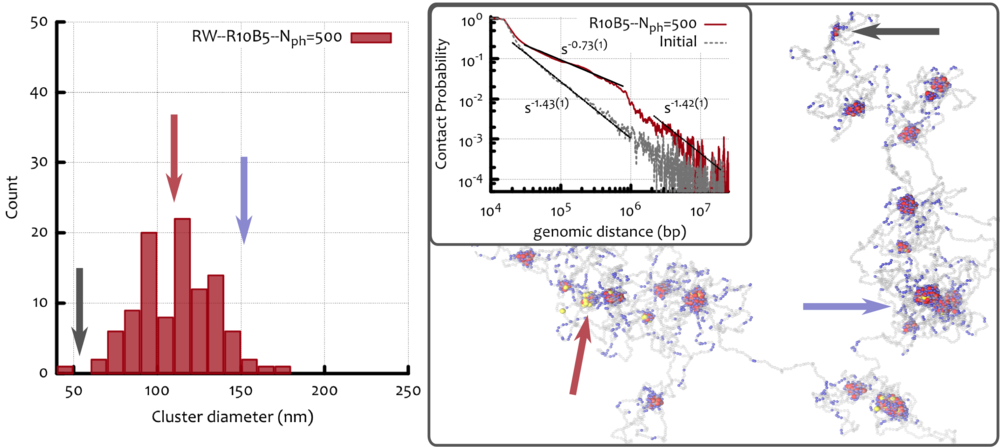}
	\caption{Cluster distribution, contact probability and a snapshot for the case in which the chromosome 3R of \emph{Drosophila} was initialised as a Random Walk (RW) and with  {$\epsilon_1= 10$} $k_BT$ and  {$\epsilon_2= 5$} $k_BT$. In this case there are clusters with a large fraction of ``polymerisation independent binding sites'' (PIBS), as highlighted by the blue arrow. In the figure, red and blue beads are respectively, strong and weak Polycomb binding sites while grey beads are neutral. Yellow beads represent PH (or PRC) bridge proteins.}
	\label{fig:SI_Clusters}
\end{figure*}

Another important aspect that has been addressed in the literature~\cite{Jost2014} is the problem of the stability of the observed TADs. Understanding whether the domains are destined to coalesce at long enough times or whether they have reached a steady state is of paramount importance in both, experiments and simulations and to understand the nature of these domains. In order to investigate this aspect we have looked at the evolution of the cluster sizes and of the contact maps for various initial conditions during the course of simulations. The simulations performed can be mapped to up to 2 or 20 minutes (depending on the viscosity $\eta_{\rm sol} \simeq 10 - 100$ cP) in real time and can therefore shed some light onto the long-time dynamics and evolution of the clusters. We computed the average size of the clusters of PH proteins by taking the average number of proteins belonging to a cluster and by assuming a spherical shape for the clusters, obtained the average diameter as $d = \langle n \rangle^{1/3} \sigma$ (we also assume the proteins to occupy a spherical volume of radius $\sigma/2$). We observed that the clusters show a slowing down at large times, in that their average size scales very weakly with time (Fig.~\ref{fig:SI_Cluster_coarsening}). In addition, the coarsening dynamics is never observed to scale faster than $t^{0.2}$ at large time. In Fig.~\ref{fig:SI_Cluster_coarsening} we report the cluster size count as a function of time for the ``polycomb-only'' (a-f) and the ``full'' (g-i) model. In Fig.~\ref{fig:SI_coarsening_Cmap_M} we also report several contact maps at different simulation time-steps for the ``full'' model and starting from the mitotic state (see above for the mitotic initialisation and Suppl. Movies M4-M6).  Fig.~\ref{fig:SI_coarsening_Cmap_M} and ~\ref{fig:SI_coarsening_Dom_M} show that the structures obtained are long lived and seem to have reached a steady state. In the main text (Fig.~8) we also show the evolution of a shorter segment between 23 and 24.5 Mbp from the initial (RW). This again  displays a remarkable long-lived structure. These figures (\ref{fig:SI_coarsening} - \ref{fig:SI_coarsening_Dom_M}) strongly suggest that the final structures that we obtain are truly stable and long lived. Whether their stability is kinetic or thermodynamic in nature is an open problem that we aim to address in the future as it has particular relevance towards the understanding of the 3D organisation of \emph{in vivo} chromatin, where the environment is strongly driven out-of-equilibrium.

%\section{Contact Probability for the Full Model}
%The contact probability for the model in which only Polycomb binding sites are considered has been shown to reproduce a probability of contact between $s$-distant segments scaling as $s^{-\gamma}$ with $\gamma \simeq 1$ within $1$ Mbp. On the other hand, $\gamma$ is shown to increase to $\gamma>1$ for non-local ($>$ 1 Mbp) contacts. This is reported Fig.~1 of the main text. In the full model, we obtain a weaker $\gamma$ for the local domains. This is consistent    get a more compact, \emph{i.e.} $\gamma \simeq 0.7$, situation for local contacts when investigated in the Full model. 
% 

\begin{figure*}[t]
	\centering
	\includegraphics[width=\textwidth]{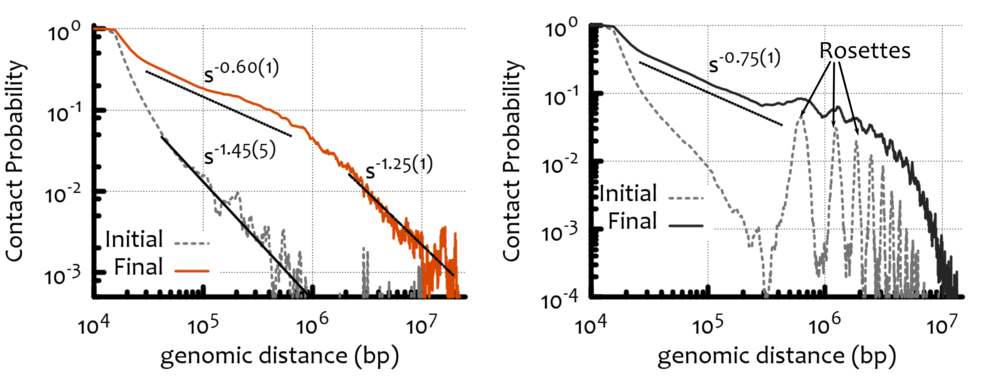}
	\caption{Contact probability for the full model starting from RW or mitotic-like configurations. One can notice that for both cases, the sub-megabase contacts decay with an exponent around $0.6-0.7$ well matching the experimentally observed one~\cite{Sexton2012}. One the other hand, beyond 1 Mbp, the contact probability is strongly affected by the choice of initial state. }
	\label{fig:SI_Cprob_full}
\end{figure*}

%\begin{figure*}[h]
%	\centering
%	\includegraphics[width=0.9\textwidth]{Figure_SI_Cprob_Full.png}
%	\caption{Contact probability computed at the start (dashed grey) and at the end (solid lines) of the ``full'' model simulations starting from RW (left) and mitotic (right) initial conditions.}
%	\label{fig:SI_Cprob_Full}
%\end{figure*} 

\end{document}